\documentclass[12pt,aps,showpacs]{revtex4}
\usepackage{graphicx}
\begin{document}
\title{Modeling full adder in Ising spin quantum 
computer \\ with 1000 qubits using quantum maps}
\author{D. I. Kamenev$^1$, G. P. Berman$^1$,
R. B. Kassman$^2$, and V. I. Tsifrinovich$^3$}
\affiliation{$^1$Theoretical Division and Center for Nonlinear Studies,
Los Alamos National Laboratory, Los Alamos, New Mexico 87545}
\affiliation{$^2$ Department of Physics, University of Illinois at
Urbana-Champaign, Urbana, Illinois 61801}
\affiliation{$^3$ IDS Department, Polytechnic University, Six
Metrotech Center, Brooklyn, New York 11201}

\begin{abstract}
The quantum adder is an essential attribute of 
a quantum computer, just as classical adder is needed for 
operation of a digital computer. We model the quantum full 
adder as a realistic complex algorithm on a large number of 
qubits in an Ising-spin quantum computer.
Our results are an important step toward effective modeling 
of the quantum modular adder which is needed for Shor's and 
other quantum algorithms. Our full adder has the following 
features: (i) The near-resonant transitions 
with small detunings are completely suppressed, 
which allows us to decrease errors by several orders of 
magnitude and to model a 1000-qubit full adder.
(We add a 1000-bit number using 2001 spins.)
(ii) We construct the full adder gates directly as 
sequences of radio-frequency pulses, rather than breaking them
down into generalized logical gates, such 
as Control-Not and one qubit gates.  
This substantially reduces the number of pulses needed
to implement the full adder. [The maximum 
number of pulses required to add one bit (F-gate) is 15].
(iii) Full adder is realized in a homogeneous spin chain.
(iv) The phase error is minimized: 
the F-gates generate approximately the same phase for 
different states of the superposition. 
(v) Modeling of the full adder is performed using quantum 
maps instead of differential equations. This allows us
to reduce the calculation time to a reasonable value.
\end{abstract}

\maketitle

\section{Introduction}
A quantum computer (QC) could efficiently solve some 
important problems using the superposition  
principle of quantum mechanics if the number of qubits in the 
register of the QC is sufficiently large. 
However, with current technologies, it is very 
difficult to implement a quantum computer with many qubits. 
It is therefore of importance to simulate and test quantum 
algorithms on digital computers. One of the main obstacles
for experimental implementation of the quantum information 
processing is decoherence caused by interaction of the quantum 
system with the environment. The second obstacle is inaccuracy 
in implementation of quantum protocol. In this paper 
we neglect these two causes of errors. 

Since an Ising spin QC based on nuclear or electron spins
is operated by radio-frequency 
({\it rf}) pulses, the wavelength of each pulse is much 
larger than the distance between qubits and much larger than
the size of the whole quantum register. That is why an Ising 
spin QC 
is characterized by the third cause of error --- nonlocality 
of interaction of the electro-magnetic waves with the qubits 
of the register. Since each {\it rf} pulse affects all 
spins in the chain, the state of each qubit of the register 
has a small probability of being changed by this pulse. For 
typical Ising spin QC 
parameters the probabilities of the unwanted excitations of 
the qubits are very small. This problem is not very important 
in a conventional NMR spectroscopy or in a QC with a small 
number of qubits because the number of pulses is relatively 
small. The operation of a QC with large number of qubits 
requires large number of pulses. Consequently errors can 
accumulate and a theoretical approach and numerical 
simulations are required to estimate this type of error.   
 
The Hilbert space of a many-qubit QC is exponentially large.
This feature creates two major problems. 
(i) Even if the number of initial states is small, the number 
of the states created by the pulses of a protocol increases 
exponentially. (ii) Due to the nonlocality of interaction of 
the {\it rf} pulses with the qubits, the direct simulation of 
the dynamics requires either the solution of large system of 
coupled differential equations for a long period of time or 
the diagonalization of large matrices. In order to overcome 
these difficulties a perturbation theory was developed in our 
previous papers~[1-9].
In this paper we apply our 
perturbation approach to simulate the dynamics of a 
full adder with 1000 qubits. 
(We suppose that there are $l=1000$ addend qubits. The 
total number of qubits including carry-over qubits is 
$L=2l+1=2001$.)

In order to solve problem (i), we formally divide 
all states of the quantum register in ``useful'' states
and ``unwanted'' states. The useful states are the states 
which realize the quantum algorithm. The unwanted states 
are the states which are created from the useful states 
by the (unwanted) action of pulses of the protocol. 
For typical QC parameters the probabilities of generation 
of the unwanted states from the useful states
are small (of the order of $\mu$, where $\mu\sim 10^{-8}$).
The probabilities of generation 
of the unwanted states from other unwanted states
are of the order of $\mu^2$, so that the latter states can be 
neglected. In spite of the fact that the total number 
of populated unwanted states increases exponentially, 
the number of unwanted states with sufficiently large 
probabilities (of the order of $\mu$) increases only linearly 
with the number of pulses. 

For solution of problem (ii), we formulate 
the quantum dynamics in terms of quantum maps. We use 
the analytic formulas from~\cite{9} of the perturbation theory
for the transition amplitudes of unwanted transitions.  
In this approach, one pulse 
of the protocol corresponds to one discrete step of the map.  

The full adder in the Ising spin QC was simulated in 
Ref.~\cite{3}. Advanced features of the quantum 
full adder presented 
in this paper are enumerated in Abstract. In Sec. II we 
consider the dynamics of the Ising spin QC. The implementation 
of full adder in the Ising spin QC is considered 
in Sec. III. Quantum protocols for 
the full adder are described in Sec. IV. The quantum map 
approach is analyzed in Sec. V. 
Our numerical results are presented in Sec. VI.  
In Sec. VII we draw our conclusions. 

\section{Ising spin quantum computer}
The Hamiltonian for the Ising spin chain placed
in an external permanent magnetic field 
and driven by rectangular {\it rf} pulses 
can be written in the form:
\begin{equation}
\label{H}
H_n=-\sum_{k=0}^{L-1}\omega_kI_k^z
-2J\sum_{k=0}^{L-2}I_k^zI_{k+1}^z
-{\Omega^{({\rm n})}\over 2}\sum_{k=0}^{L-1}
\left\{I_k^-\exp\left[-i\left(\nu_{\rm n} t+
\varphi_{\rm n}\right)\right]+
h.c.\right\}=H_0+V_{\rm n}(t).
\end{equation}
Here $\hbar=1$; $I_k^\pm=I_k^x\pm I_k^y$;
$I_k^x$, $I_k^y$, and $I_k^z$ are the components of the
operator of the $k$th spin $1/2$; $\omega_k$ is the
Larmor frequency of the $k$th spin; $J$ is the Ising 
interaction constant; $\Omega^{(n)}$ is the Rabi frequency 
(frequency of precession around the resonant transversal 
field in the rotating frame); $\nu_n$ is the frequency of 
the pulse; and $\varphi_n$ is the phase of the $n$th pulse. 
The Hamiltonian~(\ref{H}) is written for the $n$th rectangular 
{\it rf} pulse. We assume that the Larmor frequency 
of $k$th spin is $\omega_k=w_0+k\delta\omega$, so that the 
Larmor frequency difference 
$\delta\omega=\omega_{k+1}-\omega_{k}$ between the 
neighboring spins is independent of the spin number, $k$. 
Below we omit the index n which indicates the pulse number. 
The long-range dipole-dipole interaction is suppressed by 
choosing the angle between the chain and the external 
permanent magnetic field to be equal to the magic 
angle~\cite{Lopez}. Note, that the Hamiltonian (\ref{H})
allows the transitions associated with flip of only a single 
spin. 

Because of the magnetic field gradient each spin has its 
unique Larmor frequency. The selective excitations 
of the spins in the chain are performed by choosing 
the frequency of the {\it rf} pulse to be equal 
(or approximately equal) to the 
Larmor frequency of the spin which we want to excite. 	 

\subsection{Suppression of the near-resonant transitions}
\label{sec:near}
The constant Ising interaction between the spins is 
characterized by the Ising constant $J$. Due to this 
interaction, the transition frequency of $k$th spin 
depends on the states of $(k-1)$th and $(k+1)$th spins, 
so that the same spin can have different transition 
frequencies for different quantum states of a superposition. 
The $k$th spin has different transition frequencies 
for the four possible kinds of states: 
\begin{equation}
\label{states}
|\dots 0_{k-1}n_k1_{k+1}\dots\rangle,~~~~ 
|\dots 1_{k-1}n_k0_{k+1}\dots\rangle,~~~~
|\dots 0_{k-1}n_k0_{k+1}\dots\rangle,~~~~
|\dots 1_{k-1}n_k1_{k+1}\dots\rangle,
\end{equation}
where $n_k$ can be equal to 0 or 1. [The first and second 
states in Eq. (\ref{states}) have the same transition 
frequencies.] The interaction between the spins 
is necessary for implementing the conditional quantum logic 
in the QC.  

In order to ensure optimal selective excitations, a large 
magnetic field gradient (of the order 
of $10^6$ T/m~[11-13])
is required, so that 
the inequality $J\ll\delta\omega$ is satisfied. 
The differences between the transition frequencies of $k$th
spin in the states in Eq. (\ref{states}) are of order of $J$.
If one wants to flip the $k$th spin in one of the states 
in Eq. (\ref{states}) and does not want to flip the same spin 
in the other two or three states, then one must to suppress 
these unwanted transitions (near-resonant transitions) with 
the detunings (from resonance condition) of the order of $J$. 
Since $J\ll\delta\omega$, the detunings for the near-resonant 
transitions are much smaller than the detunings for the 
nonresonant transitions characterized by $\delta\omega$.  
That is why the near-resonant transitions  
in the general case generate the largest errors. 

The approach which allows one to completely suppress 
the near-resonant transitions was developed in Ref.~\cite{1}.  
In order to flip the $k$th qubit in the first two states 
in Eq. (\ref{states}) and to suppress the transitions in the 
third and fourth states, the value of the Rabi frequency 
$\Omega$ should satisfy the $2\pi K$-condition~\cite{Lopez} 
\begin{equation}
\label{Omegak}
\Omega(K)={J\over \sqrt{K^2-1/4}},\qquad K=1,2,\dots.
\end{equation}
The duration of this pulse is $\tau=\pi/\Omega(K)$ 
($\pi$-pulse) and the frequency is equal to the Larmor 
frequency $\nu^{01}_k=w_0+k\delta\omega$ of the $k$th spin. 
Here the upper indices of $\nu^{01}_k$ indicate the
states of the $(k-1)$th and $(k+1)$th spins and 
$\nu^{01}_k=\nu^{10}_k$.
The probability of the spin flip is independent of the 
phase $\varphi$ of this pulse. We denote the pulse 
with these parameters as $Q^{01}_k(\varphi)$, where the two 
upper indices indicate the states of the neighbors 
of the $k$th spin, $Q^{01}_k(\varphi)=Q^{10}_k(\varphi)$,
$k$ is the number of the spin to be 
flipped, and $\varphi$ is the phase of the pulse.  

In order to flip the $k$th qubit in the third or fourth
state in Eq. (\ref{states}) and to suppress the transitions 
in the other three states, two pulses are required~\cite{1}. 
The first 
pulse flips the $k$th spin and the second pulse removes 
the unwanted states created by the near-resonant transitions
(generated by the first pulse) from the register of the QC. 

The parameters of the first pulse required to flip 
the $k$th spin in the state 
$|\dots 0_{k-1}n_k0_{k+1}\dots\rangle$ have the following 
values~\cite{1}: the Rabi frequency is 
\begin{equation}
\label{Omega2}
\Omega_2(K_2)={2J\over \sqrt{K_2^2-1/4}},\qquad K_2=1,2,\dots;
\end{equation}
the frequency is $\nu^{00}_k=w_0+k\delta\omega+2J$; 
the time-duration of the pulse is $\tau=\pi/\Omega_2(K_2)$;
and the phase $\varphi$ of the pulse is arbitrary. 

The parameters of the second pulse required to remove 
the unwanted states created by the first pulse have the 
following values. The Rabi frequency is 
\begin{equation}
\label{Omegac}
\Omega_c(K_c)={2J\over 
\sqrt{\left({\pi K_c\over \pi+\beta}\right)^2-1}},~~~ 
K_c=1,2,\dots;
\end{equation}
the frequency is equal to the Larmor frequency, 
$w_0+k\delta\omega$, of the $k$th spin.
The time-duration, $\tau_c$, and the phase, 
$\varphi^{00}_c$, of the pulse are: 
\begin{equation}
\label{00}
\tau_c={2(\pi+\beta)\over\Omega_c(K_c)},\qquad
\varphi^{00}_c=\theta+\varphi+2Jt_0+\Theta, 
\end{equation}
where $t_0$ and $\varphi$ are, respectively, 
the beginning time and the phase of the first pulse, 
the upper index of $\varphi^{00}_c$ indicates that the 
first pulse is used to flip the $k$th spin in the 
state $|\dots 0_{k-1}n_k0_{k+1}\dots\rangle$. 
The new parameters introduced in Eqs.~(\ref{Omegac}) and  
(\ref{00}) depend only on the value of $K_2$:
\begin{equation}
\label{auxiliary}
\tan\Theta=-\sqrt{{K_2^2-1/4\over K_2^2+3/4}}
\tan\left({\pi\over 2}\sqrt{K_2^2+3/4}\right),~~~
\theta=\pi\sqrt{K_2^2-1/4},~~~
\tan\beta={1\over \sqrt{K_2^2-1/4}}\sin\Theta.
\end{equation}

The parameters of the two pulses required to flip 
the $k$th spin in the state 
$|\dots 1_{k-1}n_k1_{k+1}\dots\rangle$ have the same 
values as those for the state 
$|\dots 0_{k-1}n_k0_{k+1}\dots\rangle$, discussed above, 
except that the frequency of the first pulse, $\nu^{11}_k$, 
and the initial phase, $\varphi^{11}_c$, of the second 
pulse are
\begin{equation}
\label{11}
\nu^{11}_k=w_0+k\delta\omega-2J,\qquad
\varphi^{11}_c=-\theta+\varphi-2Jt_0-\Theta. 
\end{equation}
Below we treat the two pulses required to flip the $k$th spin 
in the state $|\dots 0_{k-1}n_k0_{k+1}\dots\rangle$ as 
one combined pulse and denote it as
$Q_k^{00}(\varphi)$. The notation $Q_k^{11}(\varphi)$ is 
used for the two pulses required to flip the $k$th spin 
in the state $|\dots 1_{k-1}n_k1_{k+1}\dots\rangle$. 

The pulses acting on the 0th qubit and $(L-1)$th qubit, 
$Q^0_k(\varphi)$ (here $k=0,L-1$ and the upper 
index indicates the state of the neighbor) have the following 
parameters:
\begin{equation}
\label{0pulse}
\nu^0_k=w_0+k\delta\omega+J,\qquad
\tau=\pi/\Omega(K),  
\end{equation}
where $\Omega(K)$ is defined by Eq.~(\ref{Omegak}) and the 
phase $\varphi$ does not affect the probability errors 
generated by this pulse. The pulse $Q^0_0(\varphi)$ 
flips the $0$th qubit in the state $|\dots0_1n_0\rangle$
($n_0=0,1$) and does not flip 
the $0$th qubit in the state $|\dots1_1n_0\rangle$.
The pulse $Q^0_{L-1}(\varphi)$ 
flips the $(L-1)$th qubit in the state 
$|n_{L-1}0_{L-2}\rangle$ and does not flip 
the $(L-1)$th qubit in the state $|n_{L-1}1_{L-2}\rangle$.
The pulse $Q^1_k(\varphi)$, $k=0,L-1$, has the frequency 
$\nu^1_k=w_0+k\delta\omega-J$ and the 
other parameters are the same as for the pulse 
$Q^0_k(\varphi)$.

\section{Quantum full adder}
The quantum full adder FA, first suggested in~\cite{adder},
adds a number $A$ to a superposition of numbers $B_i$ 
which are coded by quantum states in the register of a QC,
\begin{equation}
\label{FA}
{\rm FA(A)}\sum_{i=1}^M C_{B_i}(0)|B_i\rangle=
\sum_{i=1}^M C_{G_i}(T)|G_i\rangle e^{-iE_{G_i}T},
\end{equation}
where the state $|G_i\rangle$ is obtained as the result of 
summation of the state $|B_i\rangle$ and the number A,
the addition is performed in the interaction 
representation~\cite{interaction}, 
$E_{G_i}=\langle G_i|H_0|G_i\rangle$,
$M$ is the number of the states in the superposition, 
$T$ is the total time-duration of the full adder protocol,
$C_{B_i}(0)$ and $C_{G_i}(T)$
are the complex coefficients satisfying the 
normalization condition $\sum_{n=0}^{2^L-1}|C_n(t)|^2=1$.
For the ideal full adder these coefficients before and after 
implementation of the full adder 
are equal to each other, $C^{\rm ideal}_{G_i}(T)=C_{B_i}(0)$. 
The values of the numbers $B_i$ and $G_i$
are defined by the states 
of the spins in the spin chain. In binary notation, 
the orientation of a spin along the direction of 
the permanent magnetic field corresponds to the bit 0
and the orientation of a spin in the opposite direction 
corresponds to the bit 1. Since the quantum logic is 
implemented in the quantum computer we use the term 
``qubit'' (quantum bit) instead of the (classical) term 
``bit'' for the numbers coded through the states of the spins. 
To implement the addition, 
one should also include the carry-over qubits. 
The addition (\ref{FA}) is realized by flipping definite 
spins in definite states [see 
Eqs.~(\ref{initial1}) - (\ref{Fkkk}) below] 
in the sum in Eq.~(\ref{FA}). 
 
The quantum full adder adds numbers using the same 
rules as a classical full adder.
A classical full adder operates with an input of two 
addend bits, $a$ and $b$, and a carry-over bit , $c$,
as shown in Table I, where C and $s$ 
are, respectively, the output carry-over and sum. The latter 
can be expressed as $s=a\oplus b\oplus c$, 
where $\oplus$ is addition modulo 2. The output carry-over is 
expressed as ${\rm C}=ab\oplus ac\oplus bc$.
\begin{table}
$$
\begin{tabular}{|c|c|c||c|c|c|}\hline
\label{tab:table1}
a&b&c&$s$&C\\ \hline
~~0~~&~~0~~&~~0~~&~~0~~&~~0~~\\ \hline
0&0&1&1&0\\ \hline
0&1&0&1&0\\ \hline
1&0&0&1&0\\ \hline
0&1&1&0&1\\ \hline
1&0&1&0&1\\ \hline
1&1&0&0&1\\ \hline
1&1&1&1&1\\ \hline
\end{tabular}
$$
\caption{Table for binary addition.}
\end{table}

The quantum full adder (\ref{FA}) consists of a series of 
gates $F(a)$ (F-gates), 
where $a$ is the addend bit of the number $A$.
There are two types of gates: $F(0)$ adds the bit 0 of 
the number $A$, and $F(1)$ adds the bit 1 of the number A. 
In other words,
the value of the number A determines which protocol is 
applied to the superposition of the numbers $B_i$. The 
number of the F-gates is equal to the number of the 
addend qubits. Initially a number $B_i$ in the register is 
represented in the form
\begin{equation}
\label{initial1}
|B_i\rangle=|b^{l-1}_{L-1}0_{L-2}b^{l-2}_{L-3}0_{L-4}\dots
b^1_40_3b^0_20_10_0\rangle,
\end{equation}
where we omit the index $i$ (which indicates the state number)
in the right-hand side, the lower indices in the 
right-hand side indicate the spin number and the 
upper indices indicate the addend qubit number of the 
number $B_i$. The action of the first F-gate is defined as
\begin{equation}
\label{firstF}
F_{2,1,0}(a^0)|\dots b^1_40_3b^0_20_10_0\rangle=
|\dots b^1_40_3{\rm C}^0_2b_1^0s^0_0\rangle, 
\end{equation}
where the lower   
indices of $F_{2,1,0}(a^0)$ show that this gate acts 
on the $0$th, $1$th, and $2$nd spins. 
The output carry-over qubit of the first gate, ${\rm C}^0_2$, 
is used as an input carry-over of the next (second) F-gate,
$F_{4,3,2}(a^1)$. Using the notation introduced in Table I,
this can be written in the form $c^1_2={\rm C}^0_2$. 
The F-gate acts on intermediate qubits as
\begin{equation}
\label{Fkkk}
F_{k+1,k,k-1}(a^j)|\dots b^j_{k+1}0_kc_{k-1}^j
b_{k-2}^{j-1}\dots\rangle=
|\dots {\rm C}^j_{k+1}b^j_ks^j_{k-1}
b_{k-2}^{j-1}\dots\rangle,
\end{equation}  
where $k=2j+1$.

If the addend numbers consist of $l$ qubits, the FA(A) gate 
can be written in the form 
\begin{equation}
\label{FAA}
{\rm FA(A)}=F_{L-1,L-2,L-3}(a^{l-1})\dots 
F_{4,3,2}(a^1)F_{2,1,0}(a^0),
\end{equation}
where $|{\rm A}\rangle=|a^{l-1}a^{l-2}\dots a^1a^0\rangle$,
$L=2l+1$ is the total number of spins and 
the gate sequence should be read from right to left. 

\begin{table}[hbt]
\label{tab:table2}
\begin{center}
\begin{tabular}{|c|c|c|c|c} \cline{1-4}
 state   &  \multicolumn{3}{|c|}{acquired phase} \\ \cline{2-4}
 &   $Q_k^{01}(\varphi)$  &$Q_k^{00}(\varphi)$&$Q_k^{11}(\varphi)$&
 \\ \cline{1-4}
$|\dots 0_{k+1}0_k0_{k-1}\dots\rangle$&   $-\alpha$&
$\pi/2-\varphi+\gamma^*$&   $-\theta-\gamma$ \\

$|\dots 0_{k+1}1_k0_{k-1}\dots\rangle$&   $\alpha$&
$\pi/2+\varphi-\gamma^*$&   $\theta+\gamma$ \\

$|\dots 1_{k+1}0_k0_{k-1}\dots\rangle$&   $\pi/2-\varphi^*$&
$\pi+\theta/2+\Theta$&   $\pi-\theta/2-\Theta$ \\

$|\dots 1_{k+1}1_k0_{k-1}\dots\rangle$&   $\pi/2+\varphi^*$&
$\pi-\theta/2-\Theta$&  $\pi+\theta/2+\Theta$ \\

$|\dots 0_{k+1}0_k1_{k-1}\dots\rangle$&   $\pi/2-\varphi^*$&
$\pi+\theta/2+\Theta$&   $\pi-\theta/2-\Theta$ \\

$|\dots 0_{k+1}1_k1_{k-1}\dots\rangle$&   $\pi/2+\varphi^*$&
$\pi-\theta/2-\Theta$&   $\pi+\theta/2+\Theta$ \\

$|\dots 1_{k+1}0_k1_{k-1}\dots\rangle$&   $\alpha$&
$\theta+\gamma$&   $\pi/2-\varphi-\gamma^*$ \\

$|\dots 1_{k+1}1_k1_{k-1}\dots\rangle$&   $-\alpha$&
$-\theta-\gamma$&   $\pi/2+\varphi+\gamma^*$ \\
\cline{1-4}
\end{tabular}
\end{center}
\caption{Phases generated by the Q-pulses acting on 
intermediate qubits.
The asterisk indicates that the resonant transition from 
the state shown in the first column of the table to the 
other state, associated with the flip of the $k$th qubit, 
takes place. The phases $\theta$, $\alpha$, $\Theta$ and 
$\gamma$, are defined in Eqs.~(\ref{auxiliary}) and 
(\ref{Qphases}).}
\end{table}

\section{Quantum protocols for F-gates}
The three-qubit F-gates can be implemented using
the technique developed in Ref.~\cite{1} for one and two-qubit
gates. The pulses $Q_k^{01}(\varphi)$, 
$Q_k^{00}(\varphi)$, and $Q_k^{11}(\varphi)$ (Q-pulses), 
considered in Sec. \ref{sec:near}, flip 
the $k$th qubit only in the states with the definite 
orientations of $(k-1)$th and $(k+1)$th qubits.
[We also mention here that 
$Q_k^{01}(\varphi)=Q_k^{01}(\varphi)$.]
We say that these pulses are 
probability-corrected. Using the Q-pulses for implementation 
of the F-gates make the probability-corrected F-gates.
However, all these pulses in 
general generate different phases for different states 
of the superposition, and the differences between these phases 
for different states are not small. On the other hand 
the F-gates should generate the same phase for all states. 
In this case we say that such a protocol is phase-corrected. 
To make the protocols for F-gates phase-corrected one should 
choose the proper set of the phases $\varphi$ for the 
different Q-pulses of an F-gate. 

\begin{table}[hbt]
\label{tab:table3}
\begin{center}
\begin{tabular}{|c|c|c|c} \cline{1-3}
 state   &  \multicolumn{2}{|c|}{acquired phase} \\ \cline{2-3}
 &   $Q_0^{0}(\varphi)$  &$Q_0^{1}(\varphi)$& \\ \cline{1-3}
$|\dots 0_10_0\rangle$&$\pi/2-\varphi*$&   $-\alpha$\\
$|\dots 0_11_0\rangle$&$\pi/2+\varphi*$&   $\alpha$\\
$|\dots 1_10_0\rangle$&   $\alpha$&   $\pi/2+\varphi*$\\
$|\dots 1_11_0\rangle$&   $-\alpha$&   $\pi/2-\varphi*$\\
\cline{1-3}
\end{tabular}
\end{center}
\caption{Phases generated by the Q-pulses acting on the right 
edge ($0$th) qubit.}
\end{table}

\begin{table}[hbt]
\label{tab:table4}
\begin{center}
\begin{tabular}{|c|c|c|c} \cline{1-3}
 state   &  \multicolumn{2}{|c|}{acquired phase} \\ \cline{2-3}
 &   $Q_{L-1}^{0}(\varphi)$  &$Q_{L-1}^{1}(\varphi)$& \\ \cline{1-3}
$|0_{L-1}0_{L-2}\dots \rangle$&$\pi/2-\varphi*$&   $-\alpha$\\
$|1_{L-1}0_{L-2}\dots\rangle$&$\pi/2+\varphi*$&   $\alpha$\\
$|0_{L-1}1_{L-2}\dots\rangle$&   $\alpha$&   $\pi/2+\varphi*$\\
$|1_{L-1}1_{L-2}\dots\rangle$&   $-\alpha$&   $\pi/2-\varphi*$\\
\cline{1-3}
\end{tabular}
\end{center}
\caption{Phases generated by the Q-pulses acting on the left 
edge ($L-1$th) qubit.}
\end{table}

The phases generated by the Q-pulses acting on different 
states were calculated in Ref.~\cite{1} for the
case $K_2=K$ and $K_c=K$, when $\Omega_2=2\Omega$
and $\Omega_c\approx 2\Omega$. In order to decrease the
errors we chose smaller values for the Rabi frequencies 
$\Omega_2\approx \Omega$ and   
$\Omega_c\approx\Omega$ and take $K_2=K_c=2K$. The phases 
generated by these Q-pulses for different states are 
shown in Table II, where the phases $\theta$ and $\Theta$
are defined in Eq. (\ref{auxiliary}) and~\cite{1} 
\begin{equation}
\label{Qphases}
\alpha=\pi\sqrt{K^2-1/4},\qquad 
\gamma=\sqrt{(\pi K_c)^2-(\pi+\beta)^2}.
\end{equation}
The phases generated by the Q-pulses acting on the edge 
qubits are shown in Tables III and IV.

\begin{table}[hbt]
\label{tab:table5}
\begin{center}
\begin{tabular}{|c|c|c||c|c|c|} \cline{1-5}
pulse &  \multicolumn{2}{|c||}{$F_{k+1,k,k-1}(0)$}&\multicolumn{2}{|c|}{$F_{k+1,k,k-1}(1)$} \\ \cline{2-5}
number &~pulse~& phase $\varphi$ &~pulse~& phase $\varphi$ \\ \cline{1-5}
1 & $Q_{k}^{11}$ & $-\gamma-2\Theta$ & $Q_{k}^{11}$ & $\pi-2\gamma-4\Theta-3\theta+\alpha$ \\
2 & $Q_{k+1}^{00}$ & $7\gamma-8\Theta+\theta+14\alpha$ & $Q_{k-1}^{00}$ & $-\frac\pi 2+3\gamma-\Theta+\frac 32\theta+\alpha$ \\
3 & $Q_{k}^{00}$ & $\frac\pi 2-4\gamma+4\Theta-3\theta$ & $Q_{k-1}^{01}$ & $-\frac\pi 2-\gamma+3\Theta+\frac 12\theta$ \\
4 & $Q_{k+1}^{00}$ & $-\theta+5\alpha$ & $Q_{k-1}^{11}$ & $-\frac\pi 2-3\gamma+3\Theta-\frac 12\theta+\alpha$ \\
5 & $Q_{k-1}^{11}$ & $\pi+5\gamma-7\Theta+\frac 52\theta+5\alpha$ & $Q_{k}^{11}$ & $-4\gamma-3\theta-\alpha$ \\
6 & $Q_{k-1}^{01}$ & $\pi+2\gamma-2\Theta+\theta+2\alpha$ & $Q_{k}^{00}$ & $-\frac\pi 2-\theta$ \\
7 & $Q_{k-1}^{01}$ & $0$ & $Q_{k+1}^{00}$ & $0$ \\
8 & $Q_{k}^{00}$ & $-8\gamma+10\Theta-2\theta-11\alpha$ & $Q_{k}^{00}$ & $0$ \\
9 & $Q_{k}^{01}$ & $\pi-5\gamma+7\Theta-\frac 12\theta-7\alpha$ & $Q_{k+1}^{00}$ & $0$ \\
10 & $Q_{k}^{11}$ & $\theta-\alpha$ & $Q_{k-1}^{11}$ & $\pi-3\gamma+3\Theta-\frac 12\theta-\alpha$ \\
11 & $Q_{k-1}^{00}$ & $0$ & $Q_{k}^{01}$ & $0$ \\
12 & $Q_{k}^{00}$ & $0$ & $Q_{k}^{11}$ & $0$ \\
13 & $Q_{k}^{01}$ & $0$ & $Q_{k-1}^{00}$ & $0$ \\
14 &  & & $Q_{k}^{01}$ & $0$ \\
15 &  & & $Q_{k}^{11}$ & $0$ \\
\cline{1-5}
\end{tabular}
\end{center}
\caption{Protocols for F-gates performing addition of
an intermediate addend qubit.}
\end{table}

\begin{table}[hbt]
\label{tab:table6}
\begin{center}
\begin{tabular}{|c|c|c||c|c|c|} \cline{1-5}
pulse &  \multicolumn{2}{|c||}{$F_{k+1,k,k-1}(0)$}&\multicolumn{2}{|c|}{$F_{k+1,k,k-1}(1)$} \\ \cline{2-5}
number &~pulse~& phase $\varphi$ &~pulse~& phase $\varphi$ \\ \cline{1-5}
1 & $Q_{k}^{11}$ & $-\gamma-\theta$ & $Q_{k}^{11}$ & $\pi-2\gamma-2\Theta-\frac 52\theta-\alpha$ \\
2 & $Q_{k+1}^{0}$ & $7\gamma-8\Theta+\frac 32\theta+\alpha$ & $Q_{k-1}^{00}$ & $-\frac\pi 2+3\gamma-\Theta+\theta+\alpha$ \\
3 & $Q_{k}^{00}$ & $-\frac\pi 2-3\gamma+3\Theta$ & $Q_{k-1}^{01}$ & $-\frac\pi 2-\gamma+3\Theta$ \\
4 & $Q_{k+1}^{0}$ & $-\alpha$ & $Q_{k-1}^{11}$ & $-\frac\pi 2-3\gamma+3\Theta-\theta+\alpha$ \\
5 & $Q_{k-1}^{11}$ & $\pi+5\gamma-7\Theta$ & $Q_{k}^{11}$ & $-4\gamma+2\Theta-\theta-2\alpha$ \\
6 & $Q_{k-1}^{01}$ & $\pi+2\gamma-2\Theta+\theta+2\alpha$ & $Q_{k}^{00}$ & $\frac\pi 2-\gamma-\Theta-2\theta+\alpha$ \\
7 & $Q_{k-1}^{01}$ & $0$ & $Q_{k+1}^{0}$ & $\frac 12\theta$ \\
8 & $Q_{k}^{00}$ & $-8\gamma+10\Theta$ & $Q_{k}^{00}$ & $0$ \\
9 & $Q_{k}^{01}$ & $\pi-5\gamma+7\Theta$ & $Q_{k+1}^{0}$ & $0$ \\
10 & $Q_{k}^{11}$ & $2\alpha$ & $Q_{k-1}^{11}$ & $\pi-3\gamma+3\Theta$ \\
11 & $Q_{k-1}^{00}$ & $-\theta-\alpha$ & $Q_{k}^{01}$ & $0$ \\
12 & $Q_{k}^{00}$ & $\frac 32\theta+4\alpha$ & $Q_{k}^{11}$ & $\frac 12\theta+\alpha$ \\
13 & $Q_{k}^{01}$ & $0$ & $Q_{k-1}^{00}$ & $0$ \\
14 &  & & $Q_{k}^{01}$ & $0$ \\
15 &  & & $Q_{k}^{11}$ & $0$ \\
\cline{1-5}
\end{tabular}
\end{center}
\caption{Protocols for F-gates performing addition of the left 
addend qubit, $k=L-2$.}
\end{table}

The phase- and probability- corrected F-gates
for intermediate qubits are defined in Table V. 
The F-gates for the left addend qubit are defined in Table VI.
We must note that
the sets of phases $\varphi$ in the Tables V and VI are 
not unique since the number of linear equations 
(which is equal to 8) 
for finding the phases is smaller than the number of 
variables, $N$, in these equations~\cite{1}. 
[$N=13$ for $F(0)$ and $N=15$ for $F(1)$.]
The F-gates for the right addend qubit are 
\begin{equation}
\label{F0right}
F_{2,1,0}(0)=Q^{1}_{0}(0)
Q^{01}_{2}(0)
Q^{01}_{1}\left(-{\pi\over 2}+3\alpha\right),
\end{equation}
\begin{equation}
\label{F1right}
F_{2,1,0}(1)=Q^{0}_{0}\left({\pi\over 2}+2\alpha\right)
Q^{01}_{1}\left({\pi\over 2}+4\alpha\right).
\end{equation}
The gates (\ref{F0right}) and (\ref{F1right}) 
produce the overall phase $-3\alpha$. 

We should note that our full adder protocols are formulated 
in terms of {\it rf} pulses but not in terms of 
generalized logical gates, 
such as Control-Not and one-qubit gates. 
Our approach requires much fewer pulses which
allows us to decrease the errors significantly. 
For comparison, implementation of the F(0) gate requires 
one Control-Not gate and one Control-Control-Not gate. 
The gate F(1) also needs one more Control-Not gate and Not 
gate~\cite{3,adder}. 
Implementation of only one Control-Not gate 
requires 15 or less pulses~\cite{1}. 
In this paper we use the same number of pulses for
implementation of the complete F-gate protocols.   

\section{Quantum maps}
As discussed in Introduction,
an exact numerical simulation of the dynamics of the 
1000-qubit quantum computer is practically impossible. 
However, as shown in this Section, 
simulation of the dynamics using a perturbation approach 
is possible. The only source of errors in our model of quantum 
computer are the nonresonant transitions. We use
analytical formulas for the amplitudes of the 
nonresonant transitions to formulate the dynamics in terms of 
quantum maps. One step of the map corresponds to a single 
Q-pulse of the protocol. Each map consists of the following 
four successive steps: (a) Each useful 
state of a superposition is mapped to another useful state 
according to the algorithm. 
If the transition is resonant the useful state is changed, 
if the transition is near-resonant, the useful state 
is not changed. The probability amplitudes of the 
useful states are not changed and are assumed to be real 
since the phases of the useful states
are not important for the calculation of the probability 
errors: the unwanted states, generated from the useful states 
have random phases (see below). (b) Each existing unwanted 
state is mapped to another unwanted state according to 
the algorithm due to resonant transitions in the same 
manner as the useful states. 
(After generation, the unwanted states evolve as 
the useful states under the resonant action of the {\it rf} 
pulses.) The existing phase of each 
unwanted state is not changed. Actually, this phase is 
not important because the phases of
the amplitudes of the nonresonant transitions, 
which contribute to this unwanted state are assumed to be 
random. (c) Each useful state is mapped to a finite number 
(which is equal or less than $2L$) of unwanted states 
(nonresonant transitions). The details of this step 
are described in the rest of this Section. 
(d) The total probability of unwanted states is 
calculated and the probability amplitudes of the useful states 
are multiplied by a factor, close to unity, 
to satisfy the normalization condition.
The total probability error is calculated as a sum 
of the probabilities of all unwanted states. 

\subsection{Probability amplitudes for the 
nonresonant transitions}
The probability amplitudes for the nonresonant transitions 
were calculated in Ref.~\cite{9}. If the state $|i\rangle$
is initially populated, $C_i(t_0)=1$,
then after application of 
one rectangular pulse with duration $\tau$ and 
frequency $\nu$, resonant or near-resonant with the 
transition frequency of the $k$th spin, the probability 
amplitudes for the nonresonant transitions 
associated with flip of $k'$th spin ($k'\ne k$) are   
$$
C_m(t_0+\tau)=-{\Omega\over 2D}
\left\{\cos\left({\Lambda \tau\over 2}\right)-
\left[\cos\left({\lambda \tau\over 2}\right)+
i\eta\frac\delta\lambda
\sin\left({\lambda \tau\over 2}\right)\right]
e^{iD\tau}\right\}\times
$$
$$
\exp\left\{i\left[\left(D+{\eta\delta-\Delta\over 2}\right)t_0
-{\Delta\over 2}\tau-
\left(\sigma-\frac 12(1-\eta)\right)\varphi
\right]\right\}
$$
\begin{equation}
\label{dynamics}
C_n(\tau)=-i{\Omega\over 2D}
\left\{\frac\Omega\Lambda
\sin\left({\Lambda \tau\over 2}\right)-
\frac\Omega\lambda
\sin\left({\lambda \tau\over 2}\right)
e^{iD\tau}\right\}\times
\end{equation}
$$
\exp\left\{i\left[\left(D+{\eta\delta+\Delta\over 2}\right)t_0
+{\Delta\over 2}\tau-
\left(\sigma+\frac 12(1+\eta)\right)\varphi
\right]\right\}.
$$
Here the state $|m\rangle$ is related to the initial state 
$|i\rangle$ by flip of $k'$th spin; 
the state $|n\rangle$ is related to the initial state 
by flips of $k$th and $k'$th spin; 
$\sigma=1$ if the $k'$th spin of the state $|i\rangle$ 
is in the state $n_{k'}=0$ and $\sigma=-1$ if the $k'$th 
spin of the state $|i\rangle$ is in the state $n_{k'}=1$. 
The other parameters are: 
\begin{equation}
{\rm if}~n_k=0~{\rm then}~ 
\left\{
\begin{array}{l}
\eta=1 \\
\delta=E_j-E_i-\nu \\
\Delta=E_n-E_m-\nu \\
D=E_m-E_i-\sigma\nu \\
+(\Delta-\delta)/2 
\end{array}\right.,~~~~
{\rm if}~n_k=1~{\rm then}~ 
\left\{
\begin{array}{l}
\eta=-1 \\
\delta=E_i-E_j-\nu \\
\Delta=E_m-E_n-\nu \\
D=E_n-E_j-\sigma\nu \\
+(\Delta-\delta)/2
\end{array}\right.,
\end{equation}
$$
\lambda=\sqrt{\Omega^2+\delta^2},\qquad 
\Lambda=\sqrt{\Omega^2+\Delta^2}.
$$ 
Here $\delta$ is the detuning for the resonance or 
near-resonance transition $|i\rangle\rightarrow |j\rangle$
associated with flip of the $k$th spin;
$\Delta$ is the detuning for the resonance or near-resonance 
transition between the unwanted states
$|m\rangle\rightarrow |n\rangle$;
$n_k$ is the state of the $k$th spin of the 
initial state $|i\rangle$. 
The transition amplitudes for the non-resonant transitions in 
Eq. (\ref{dynamics}) are 
characterized by the large detuning 
$|D|\approx|k-k'|\delta\omega$~\cite{1,2,4,9}, 
which is approximately equal to the distance between 
the $k$th and $k'$th spins measured in the frequency 
units.

\subsection{Random phases for the unwanted states}
The analytical formula (\ref{dynamics}) and 
for the transition 
amplitudes of the nonresonant transitions are derived for 
one pulse only. We now discuss how to use 
these equations for simulations of quantum protocols
which consist of many rectangular pulses with different 
parameters. First assume that initially only one state 
is populated in the register of our computer. 
The first pulse of the protocol creates approximately 
$2L$ unwanted states as a result of the nonresonant 
transitions. The amplitudes of all nonresonant 
transitions, $C_m$, can be calculated using 
Eq. (\ref{dynamics}) without integration of the 
Schr\"odinger equation. The probabilities of populations
of these states are given by the square moduli $P_m=|C_m|^2$
and the phases are equal to the arguments of $C_m$,
$\phi_m={\rm arg}C_m$. 
Since in Eq.~(\ref{dynamics}) 
$t_0\ge\tau\sim \pi/\Omega$ and $D/\Omega\sim 10^4\gg 1$ 
the values of $\phi_m$ oscillate rapidly as functions of 
$t_0$ and can be assumed to be random. In our 
simulations we assume random values for $\phi_m$ 
in spite of the fact that the exact values are known. 
[$\phi_m$ are equal to the arguments of the exponents 
containing $Dt_0$ in Eq. (\ref{dynamics}).]
We do this in order to make our method applicable to 
other possible quantum computer models where the 
probabilities of the nonresonant transitions can 
be estimated analytically, but the phases can not. On the 
other hand, as was shown above, the information 
about the exact values of these phases is not important.
 
\subsection{Linear accumulation of probability errors}
When the subsequent pulses of the protocol contribute 
to the same unwanted state, the corresponding probability 
amplitudes should be added to each other. The
addition of complex numbers with the random phases 
can increase or decrease
the modulus of the sum, 
depending on the difference between the phases 
of the addend numbers. Consequently, it is 
possible, in principle, for the 
{\it rf} pulse to decrease the total error. 
The result of the action of many pulses on one unwanted 
state can be presented as a multiple algebraic addition 
of complex numbers with random phases and the change 
of the probability amplitude of each unwanted state 
can be described by the random walk model in 
two-dimensional (complex) plane. Due to this model, 
the probability amplitudes of the unwanted states 
grow as $|C_m|\sim\sqrt N$, where $N$ is the number of pulses,
and the total probability error is proportional to $N$.

\section{Numerical results}
First, we simulate the full adder protocol 
for a small number of qubits using an exact numerical 
solution~\cite{2,4} 
in order to calculate the phase errors and to test the quantum 
map approach. For the relation between the numerical and 
physical parameters see Refs.~\cite{1,Lopez}.

\subsection{Phase error}
The phase error, $P_{\rm ph}$, is defined as 
\begin{equation}
\label{Ph}
P_{\rm ph}={\rm max}_i|\delta\phi_{G_i}-\overline{\delta\phi}|,
\end{equation}
where 
\begin{equation}
\label{Ph1}
\delta\phi_{G_i}=\phi_{G_i}(T)-\phi_{B_i}(0),~~~ 
\phi_{B_i}(0)=
{\rm arctan}{{\rm Im}C_{B_i}(0)\over{\rm Re}C_{B_i}(0)},~~~
\phi_{G_i}(T)=
{\rm arctan}{{\rm Im}C_{G_i}(T)\over{\rm Re}C_{G_i}(T)},
\end{equation}
where Im and Re stands, respectively, for the imaginary 
and real parts.
The common phase
\begin{equation}
\label{average_phi}
\overline{\delta\phi}={1\over M}\sum_{i=1}^M
[\phi_{G_i}(T)-\phi_{B_i}(0)|
\end{equation}
is close to the value $-3\alpha$ given by 
the protocol. The coefficients $C_{B_i}(0)$, $C_{G_i}(T)$, 
and $M$ are introduced in Eq. (\ref{FA}). The phase error 
in the FA(A) gate caused 
by nonresonant transitions is shown in Fig. 1  
for different values of the addend number A for
$l=5$ addend qubits. (The total number of qubits is $L=11$.) 
Initially, the superposition contained $M=4$ numbers:
7, 12, 16, and 27
with randomly chosen complex coefficients $C_{B_i}(0)$
satisfying the normalization condition 
$\sum_{i=1}^M|C_{B_i}(0)|^2=1$. 
The phase error in Fig.~1 is of the order or less 
than 1\% of $\pi$. This error can be decreased 
by increasing $\delta\omega$
or decreasing $\Omega$~\cite{1}. The phase error increases
as the number of qubits increases~\cite{1}. 

\begin{figure}
\centerline{\includegraphics[width=8cm,height=8cm]{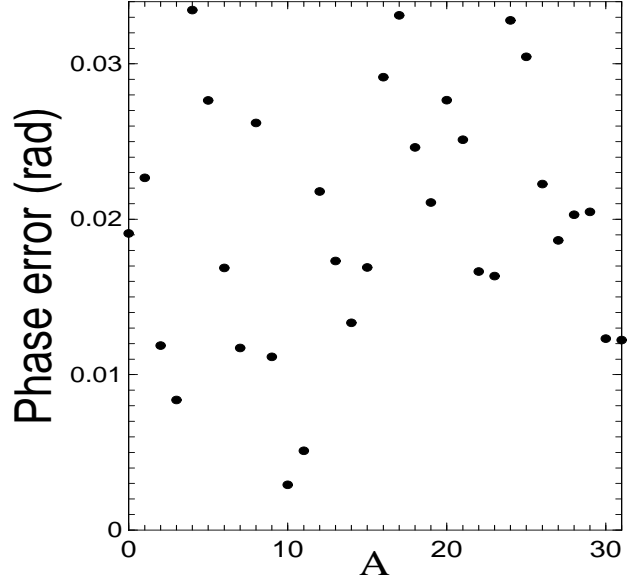}}
\vspace{-5mm}
\caption{Phase error after implementation of the full adder 
for different values of the addend number A. 
$K=8$ [$\Omega\approx J/8$, see Eq. (\ref{Omegak})],  
$\delta\omega/\Omega=10^4$ ($\delta\omega\approx 1252 J$),
$l=5$ ($L=11$).}
\label{fig:1}
\end{figure}

\begin{figure}
\centerline{\includegraphics[width=8cm,height=8cm]{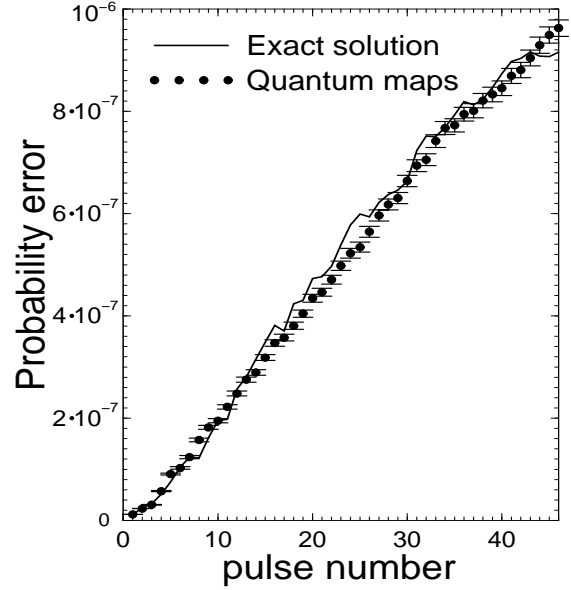}}
\vspace{-5mm}
\caption{The probability error as a function of the number 
of Q-pulses during implementation of the full adder 
obtained using exact solution and using quantum maps. 
$K=100$ ($\Omega\approx 0.01$), $\delta\omega=100$, 
$l=4$ ($L=9$). 
The data obtained using the quantum maps 
are averaged over 100 realization with different sets 
of the random phases.} 
\label{fig:2}
\end{figure}

\subsection{Test of the quantum map approach}
Since we used random phases for probability amplitudes 
generated by the unwanted transitions instead of 
their actual phases, we tested our approach by 
comparison with the exact numerical solution. 
In Fig. 2 we compare the probability errors calculated 
using quantum maps with the probability errors 
computed using the exact solution for 
$L=9$ qubits. The initial superposition contained
the numbers 2, 5, 11, 12 with randomly chosen 
normalized coefficients $C_{B_i}(0)$ and A=6.
The probability error for the exact case is defined as 
\begin{equation}
\label{pr_error}
P=\sum_{i=1}^M||C_{G_i}(T)|^2-|C_{B_i}(0)|^2|,
\end{equation}
while the probability error using quantum maps is defined 
as the sum of the probabilities of all unwanted states. 
As follows from Fig. 2, 
there is good correspondence between the data obtained 
using the exact numerical solution and the results obtained
using the quantum maps. 
We observed a similar correspondence 
for the other initial conditions and other addend numbers A.    

\begin{figure}
\centerline{\includegraphics[width=8cm,height=8cm]{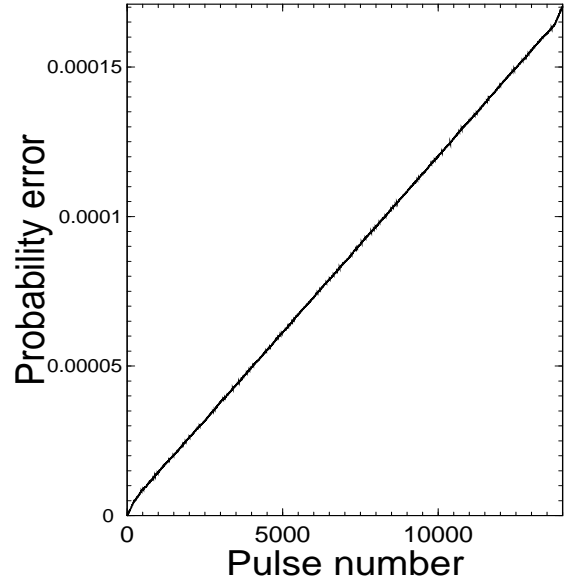}}
\vspace{-5mm}
\caption{The probability error as a function of the number 
of Q-pulses during implementation of the full adder 
obtained using quantum maps. 
$K=100$ ($\Omega\approx 0.01$), $\delta\omega=100$.
Number of addend qubits is $l=1000$ and the total number 
of qubits is $L=2001$.} 
\label{fig:3}
\end{figure}

\subsection{Modeling the full adder with 1000 addend qubits}
The growth of error with increasing the number of pulses
is shown in Fig. 3. This figure is obtained using quantum 
maps with 20 randomly chosen initial numbers, $B_i$, with 
randomly chosen normalized coefficients, $C_{B_i}(0)$. 
The data are averaged over 20 different realizations
with different sets of the random phases. The error bars 
are of the order of or less than the line width. We also 
modeled the full adder for superposition of 1 and 100
states and obtained the same curve as in Fig. 3, so 
that the probability error appears to be 
independent of the number 
of states in the superposition. During modeling 
only the unwanted states with probabilities 
$\xi\ge 0.001(\Omega/\delta\omega)^2/M$ were taken into 
consideration. All other unwanted states with the smaller
probabilities were neglected. Decreasing the value 
of $\xi$ did not significantly affect the 
value of the probability error. As discussed above, 
the probability error in Fig. 3 grows linearly
and is approximately equal to $N(\Omega/\delta\omega)^2$,
where $N$ is the number of pulses. In Fig. 4 we plot 
the number of unwanted states generated by one useful state 
as a function of the number of Q-pulses. If the initial 
superposition contains $M$ useful states the number of 
unwanted states must be multiplied by $M$. 

\begin{figure}
\centerline{\includegraphics[width=8cm,height=8cm]{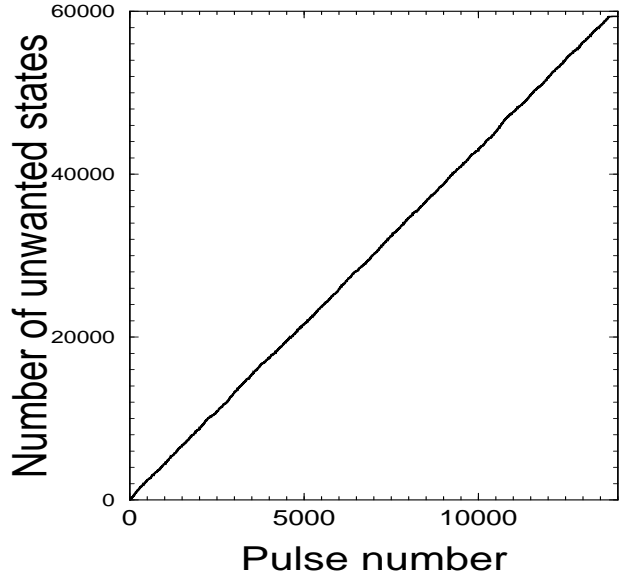}}
\vspace{-5mm}
\caption{The growth of the number of unwanted states 
generated by one useful state as a function of 
the number of Q-pulses implementing  
the full adder protocol. The parameters are the same 
as in Fig. 3. The data are averaged over 100
quantum map realizations.}
\label{fig:4}
\end{figure}

\section{Conclusion}
We have successfully demonstrated in this paper that a 
very efficient mapping procedure can closely approximate 
exact quantum dynamics of quantum computer.
Our quantum map approach can also be generalized 
for use in other models, 
including models with time-dependent Hamiltonians. 
(Our Hamiltonian is time-independent in the rotating frame.)
For a model with nearest neighbor constant 
interaction between the qubits, the result 
of the action of a pulse depends 
on the frequency difference between $k$th and $k'$th
spin and on orientations of
the following spins: $k$th spin with resonant or near-resonant 
frequency; $(k-1)$th and $(k+1)$th spins; 
$k'$th spin with nonresonant frequency, whose 
flip creates an error; $(k'-1)$th and $(k'+1)$th spins. 
The number of these possible configurations is relatively 
small even for a computer with large number of qubits. 
Even in the case when the analytic solution 
is unknown, the results of the action of a single pulse 
on the finite number of possible spin configurations
enumerated above can be calculated numerically to 
define the moduli for the transition 
amplitudes of one step of the discrete map. 
As shown in this paper the phases of the 
generated unwanted states can be 
assumed to be random. After identifying the maps, one can 
use them to simulate a whole quantum protocol. 

The quantum map method is similar to dynamical simulations 
of quantum systems with time-periodic Hamiltonians 
using the Floquet theorem~\cite{Reichl,Reichl1,chaos}. 
In the latter approach one calculates the dynamics 
for one period of external field, 
${\cal T}$. Then one builds the evolution operator for one 
period, and then the evolution operator for one period 
can be used to find the state of the system at time 
$m{\cal T}$, $m=2,3\dots$, without integration of the 
Schr\"odinger equation for this time. The Floquet theorem
is especially useful for finding the asymptotic behavior 
of the system, for $m\rightarrow\infty$. Similarly, the 
quantum map approach is most useful when a quantum 
protocol contains a large number of pulses and when 
the number of qubits in the register of a quantum computer 
is large.  

\begin{acknowledgments}
We are grateful to G. D. Doolen for useful discussions.
This work was supported by the Department of Energy (DOE) 
under Contract No. W-7405-ENG-36, by the National Security 
Agency (NSA), and by the Advanced Research and Development 
Activity (ARDA). RBK acknowledges partial support from the 
National Science Foundation under grant NSF-EIA-01-21568, 
and from the Center for Nonlinear Studies, Los Alamos
National Laboratory.

\end{acknowledgments}

{}
\end{document}